\documentclass[12pt]{article}
\usepackage{epsfig}
{\def\mtiny{\vrule width 0pt}
\def\mrm#1{\mathrm{#1}}
\def\DZ{\relax\ifmmode{D^0}\else{$\mrm{D}^{\mrm{0}}$}\fi}
\def\DONE{\relax\ifmmode{D_1}\else{$\mrm{D}_{\mrm{1}}$}\fi}
\def\DTWO{\relax\ifmmode{D_2}\else{$\mrm{D}_{\mrm{2}}$}\fi}
\def\KZ{\relax\ifmmode{K^0}\else{$\mrm{K}^{\mrm{0}}$}\fi}
\def\KSHO{\relax\ifmmode{K_{\rm S}}\else{$\mrm{K}_{\mrm{S}}$}\fi}
\def\KLON{\relax\ifmmode{K_{\rm L}}\else{$\mrm{K}_{\mrm{L}}$}\fi}
\def\BZ{\relax\ifmmode{B^0_d}\else{$\mrm{B}^{\mrm{0}_d}$}\fi}\def\BZp{\relax\ifmmode{B^0}\else{$\mrm{B}^{\mrm{0}}$}\fi}
\def\BZS{\relax\ifmmode{B^0_s}\else{$\mrm{B}^{\mrm{0}_s}$}\fi}
\def\DZS{\relax\ifmmode{D^{*+}}\else{$\mrm{D}^{\mrm{*+}}$}\fi}
\def\DZB{\relax\ifmmode{\overline{D}\mtiny^0}
        \else{$\overline{\mrm{D}}\mtiny^{\mrm{0}}$}\fi}\def\KZB{\relax\ifmmode{\overline{K}\mtiny^0}
        \else{$\overline{\mrm{K}}\mtiny^{\mrm{0}}$}\fi}
\def\BZB{\relax\ifmmode{\overline{B}\mtiny^0_d}
        \else{$\overline{\mrm{B}}\mtiny^{\mrm{0}_d}$}\fi}
\def\BZBp{\relax\ifmmode{\overline{B}\mtiny^0}
        \else{$\overline{\mrm{B}}\mtiny^{\mrm{0}}$}\fi}
\def\BZBS{\relax\ifmmode{\overline{B}\mtiny^0_s}
        \else{$\overline{\mrm{B}}\mtiny^{\mrm{0}_s}$}\fi}
\def\DZC{\relax\ifmmode{\overline{D}\mtiny^0}
        \else{$\overline{\mrm{D}}\mtiny^{\mrm{0}}$}\fi}

\begin{document}

\title{\vspace{-1in}
\begin{flushright}
\begin{minipage}{1.5in}
	\normalsize
	UCSB HEP 99-09
\end{minipage}
\end{flushright}
\vspace{0.75in}
Search for $\DZ\!-\!\DZB$ Mixing}
\author{Harry N. Nelson\\
Physics Department\\
University of California\\
Santa Barbara, CA 93106-9530}
\date{\hfil\vspace{-1.2cm}}
\maketitle

\begin{abstract} 
We report on a search for $\DZ\!-\!\DZB$ mixing made by
a study of the `\hbox{wrong-sign}' process $\DZ\!\to\!K^+\pi^-$.
The data come from an
integrated luminosity of $e^+e^-$
collisions at $\sqrt{s}\approx10\,$GeV
consisting of 9.0~fb$^{-1}$,
recorded with the CLEO-II.V detector.
We measure the time-integrated rate of the `\hbox{wrong-sign}' process
$\DZ\!\to\!K^+\pi^-$, relative to that of the Cabibbo-favored
process $\DZB\!\to\!K^+\pi^-$, to be
$R_{\rm ws}=(0.34\pm0.07\pm0.06)\%$.  
We study that rate as a function
of the decay time of the $\DZ$, to distinguish the
rate of direct doubly-Cabibbo-suppressed decay from
$\DZ\!-\!\DZB$ mixing.
The amplitudes that describe
$\DZ\!-\!\DZB$ mixing, $x^\prime$ and $y^\prime$, are consistent
with zero.  The one-dimensional limits, at the 95\% C.L., that
we determine are $(1/2)x^{\prime2}<0.05\%$, and
$-5.9\%\!<\!y^\prime\!<\!0.3\%$.  All results are preliminary.
\end{abstract}

Studies of the evolution of a $\KZ$ or $\BZ$ into the respective
anti-particle, a $\KZB$ or $\BZB$,
have guided the form and content
of the Standard Model, and permitted
useful estimates of the 
masses of the charm and top quark masses, prior to 
direct observation of those quarks.
In this paper, we present
the results of a search for the evolution of the $\DZ$ into
the $\DZB$, where the principal motivation is to glimpse
new physics outside the Standard Model, prior to direct observation
of that physics at the high energy frontier.

A $\DZ$ can evolve into a $\DZB$ through on-shell intermediate
states, such as $K^+K^-$ with $m_{K^+K^-}\!=\!m_{\DZ}$, or through
off-shell intermediate states, such those that might be present
due to `new physics'.  We denote the amplitude through the former (latter)
states by $-iy$ ($x$), in units of $\Gamma_{\DZ}/2$~\cite{ampli}, and
we neglect all types of CP violation. 

For comparison,
in the $\KZ\!-\!\KZB$ system the analogous $y$ and $|x|$ are 
both near unity~\cite{RPP98,xyk}.
The prediction that an initial $\KZ$ will
decay with two lifetimes, as described by the non-zero $y$,
was famously made by Gell-Mann and Pais~\cite{gmp}.
In the $\BZ\!-\!\BZB$ system theory firmly predicts that $y$ is negligible, 
and experiments have not yet sought $y$ out.
For the $\DZ\!-\!\DZB$ system, $y$ is likely to receive significant
contributions from the Standard Model~\cite{gp}.  It may even be that
$y$ dominates the $\DZ\!\to\!\DZB$ amplitude.

It has been $x$, for both the $\KZ\!-\!\KZB$ and $\BZ\!-\!\BZB$ 
systems, that has provided information about the charm and 
top quark masses.  A report
of the first measurement in 1961 of $|x|$ 
for the $\KZ\!-\!\KZB$ system
noted `We cannot compare our experimental value for $|x|$ with any
theoretical calculation' \cite{goodetal}.  
By 1974, theory had caught up, and
exploited $|x|$ for the $\KZ\!-\!\KZB$ system
to predict the charm quark mass~\cite{glr}, just before that
quark was discovered.  The tiny, CP violating
${\rm Im}(x)$ for the $\KZ$ is sensitive to the value of the top quark mass.
The large value for the $\BZ$ of $|x|$ (now $0.73\pm0.03$~\cite{RPP98})
indicated that the top quark is very massive~\cite{rosner87}, in 
distinction to contemporaneous data from the high energy
frontier~\cite{ua1top}.

Many predictions for $x$ in the $\DZ\!\to\!\DZB$ amplitude have
been made~\cite{hnncomp}.  The Standard Model contributions are suppressed
down to at least $|x|\approx\tan^2\theta_C\approx0.05$ because
$\DZ$ decay is Cabibbo-favored; the
GIM~\cite{gimktwz} cancellation could 
suppress $|x|$ down to $10^{-6}-0.01$.
Many non-Standard Models, particularly those that address
patterns of quark flavor, predict $|x|\approx0.01$ or greater.  Contributions
to $x$ at this level can result from
the presence of new particles with masses as high as 100~TeV~\cite{lns}.
Decisive signatures of such particles might include
$|y|\ll|x|$, or CP-violating interference between
a substantial imaginary component of $x$ and either $y$, or
a direct decay amplitude.  In order to accurately assess the origin
of a $\DZ\!-\!\DZB$ mixing signal, the effects described by $y$ must be
distinguished from those that are described by $x$.

We report here on a study of the process
$\DZ\!\to\!K^+\pi^-$. We use the charge of the `slow' 
pion, $\pi_s^+$, from the decay $D^{*+}\!\to\!\DZ\pi^+_s$ to deduce
production of the $\DZ$, and then we seek  the rare
`\hbox{wrong-sign}' $K^+\pi^-$ final state (WS), in addition to
the more frequent  `\hbox{right-sign}' final state, $K^-\pi^+$ (RS).
The \hbox{wrong-sign} process, $\DZ\!\to\!K^+\pi^-$, can proceed either
through direct doubly-Cabibbo-suppressed decay (DCSD),
or through mixing followed by the Cabibbo-favored
decay (CFD), $\DZ\!\to\!\DZB\!\to\!K^+\pi^-$.  Both processes contribute
to the time integrated `\hbox{wrong-sign}' rate, $R_{\rm ws}$:
\begin{displaymath}
R_{\rm ws}=\displaystyle{\Gamma(\DZ\!\to\!K^+\pi^-)\over\Gamma(\DZB\!\to\!K^+\pi^-)}.
\end{displaymath}

To disentangle the two processes that could contribute to
$\DZ\!\to\!K^+\pi^-$, we study the distribution of \hbox{wrong-sign}
final states as a function of the proper decay time, $t$, 
of the $\DZ$.  The proper decay time is in units of 
the mean $\DZ$ lifetime, $\tau_{\DZ}=415\pm4\,$fs~\cite{RPP98}.
The differential `wrong-sign' rate, relative to
$\Gamma(\DZB\!\to\!K^+\pi^-)$, is $r_{\rm ws}(t)$~\cite{timev,bsn}:
\begin{equation}
r_{\rm ws}(t)\equiv[R_D + \sqrt{R_D} y^\prime t + 
         \displaystyle{1\over4}(x^{\prime2}\!+\!y^{\prime2})t^2]e^{-t},
\label{eq:rws}
\end{equation}
where the modified mixing amplitudes $x^{\prime}$ and $y^{\prime}$ in
Eqn.~\ref{eq:rws} are given by:
\begin{displaymath}
\begin{array}{rcl}
 y^\prime&\equiv&y\cos\delta\!-\!x\sin\delta\\
 x^\prime&\equiv&x\cos\delta\!+\!y\sin\delta
\end{array},
\end{displaymath}
and $\delta$ is a possible strong phase between DCSD and CFD
amplitudes:
\begin{displaymath}
-\sqrt{R_D}e^{-i\delta}\equiv
{\langle K^+\pi^-|T|\DZ\rangle\over
\langle K^+\pi^-|T|\DZB\rangle}.
\end{displaymath}
The symbol $R_D$ represents the DCSD rate, relative to the CFD rate.
There are plausible theoretical arguments 
that $\delta$ is small~\cite{wfsp,brpa}.  If
both $\DZ$ and $\DZB$ exclusively populate the $I\!=\!1/2$
amplitude of $K^+\pi^-$, then $\delta$ would be zero.  For
the CFD, $\DZB\to K^+\pi^-$, the $I\!=\!3/2$ amplitude is indeed
disfavored, with $|A_{3/2}/A_{1/2}|=0.27\pm0.02$~\cite{withe}.  A
non-zero $\delta$ could develop if $\DZ\!\to\!K^+\pi^-$ populates
the $I\!=\!3/2$ amplitude differently than the CFD does.  Crudely,
one might guess $|\delta|\!<\,\sim\!|A_{3/2}/A_{1/2}|\!\sim15^\circ$.
The size of $\delta$ could be settled by measurements
of the DCSD contributions to $D^+\!\to\!K^+\pi^0$, $D^+\!\to\!\KLON\pi^+$,
and $\DZ\!\to\!\KLON\pi^0$, which are now feasible
with our data set.

For decays to wrong-sign final states other than $K^+\pi^-$, such
as $K^+\pi^-\pi^0$, or $K^+\pi^-\pi^+\pi^-$, there will be distinct
strong phases.  Moreover, the broad resonances that
mediate those multibody hadronic decays, 
such as the $K^*$, $\rho$, etc., modulate
those phases as a function of position on the Dalitz plot.  Thus,
multibody hadronic wrong-sign decays might afford an opportunity to
distinguish $x$ and $y$ from $x^{\prime}$ and $y^{\prime}$.

An important aspect of the expression Eqn.~\ref{eq:rws} for
the two-body decay $\DZ\!\to\!K^+\pi^-$ is that
the dependence on $y^{\prime}$ and $x^{\prime}$ is distinguishable
due to the interference with the direct decay amplitude, which
induces the term linear in $t$.
Such behavior is complementary
to the differential decay rate to CP eigenstates such as
$\DZ\!\to\!K^+K^-$, which is sensitive to $y$ alone, or that of
$\DZ\!\to\!K^+\ell^-\overline{\nu}_{\ell}$, which is sensitive to
$r_{\rm mix}\equiv(x^2\!+\!y^2)/2=(x^{\prime2}\!+\!y^{\prime2})/2$ alone.

Our data was accumulated between Jan.~1996 and
Feb.~1999 from an integrated luminosity of $e^+e^-$
collisions at $\sqrt{s}\approx10\,$GeV
consisting of 9.0~fb$^{-1}$, provided by 
the Cornell Electron Storage Rings (CESR).
The data were taken with CLEO-II multipurpose 
detector~\cite{ctwo}, upgraded in
1995 when a silicon vertex detector (SVX) was installed~\cite{csvx}
and the drift chamber gas was changed from argon-ethane to helium-propane.
The upgraded configuration is named CLEO-II.V, where the `V' is short
for Vertex.

We reconstruct candidates for the decay sequences $D^{*+}\!\to\!\pi^+_s\DZ$,
followed by either $\DZ\!\to\!K^+\pi^-$ (\hbox{wrong-sign}) or
$\DZ\!\to\!K^-\pi^+$ (\hbox{right-sign}). The sign of the slow charged
pion tags the charm state at $t=0$ as either $\DZ$ or
$\DZB$.  The broad features of the
reconstruction are similar to those employed in the recent
CLEO-II.V measurement of the $D$ meson lifetimes~\cite{dlife}.
The following points are important in understanding how we have
improved our sensitivity to $\DZ\!\to\!K^+\pi^-$, relative to
earlier CLEO work~\cite{tl,leppho}:
\begin{enumerate}
\item The SVX allows substantially more precise measurement of
charged particle trajectories in the dimension parallel to the
colliding beam axis.  When combined with improvements in track-fitting,
the CLEO-II.V resolution for $Q=M_{\pi_s K\pi}\!-\!M$, where $M$ is
the mass of the $K^+\pi^-$ system,
is $\sigma_Q=190\pm6\,$KeV, compared with the earlier value
of $\sigma_Q=750\,$KeV~\cite{tl}.
\item The use of helium-propane, in addition to improvements in
track-fitting, have reduced the CLEO-II.V resolution for
$M$ to $\sigma_M=6.4\pm0.1\,$MeV, compared with the earlier
value of $\sigma_M=11\,$MeV~\cite{tl}.
\item Improved mass resolution, as well as rejections based
on the momentum asymmetry of the two charged tracks, allow clean separation
of the signal $\DZ\!\to\!K^+\pi^-$ from $\DZ\!\to\!K^+K^-$, 
$\DZ\!\to\!\pi^+\pi^-$, and $\DZ\!\to\!K^-\pi^+$, and from
multibody decays of the $\DZ$, at a cost of about 35\% of 
the signal acceptance. We use the modest $\pi^+/K^+$ separation provided by
measurement of $dE/dx$ in the CLEO-II.V drift chamber primarily
for  systematic studies.  Addition of a new device 
with perfect $\pi^+/K^+$ identification and acceptance 
might enable the recovery of the 35\% acceptance loss.
\end{enumerate}
\begin{figure}[htpb]
\begin{center}
\epsfig{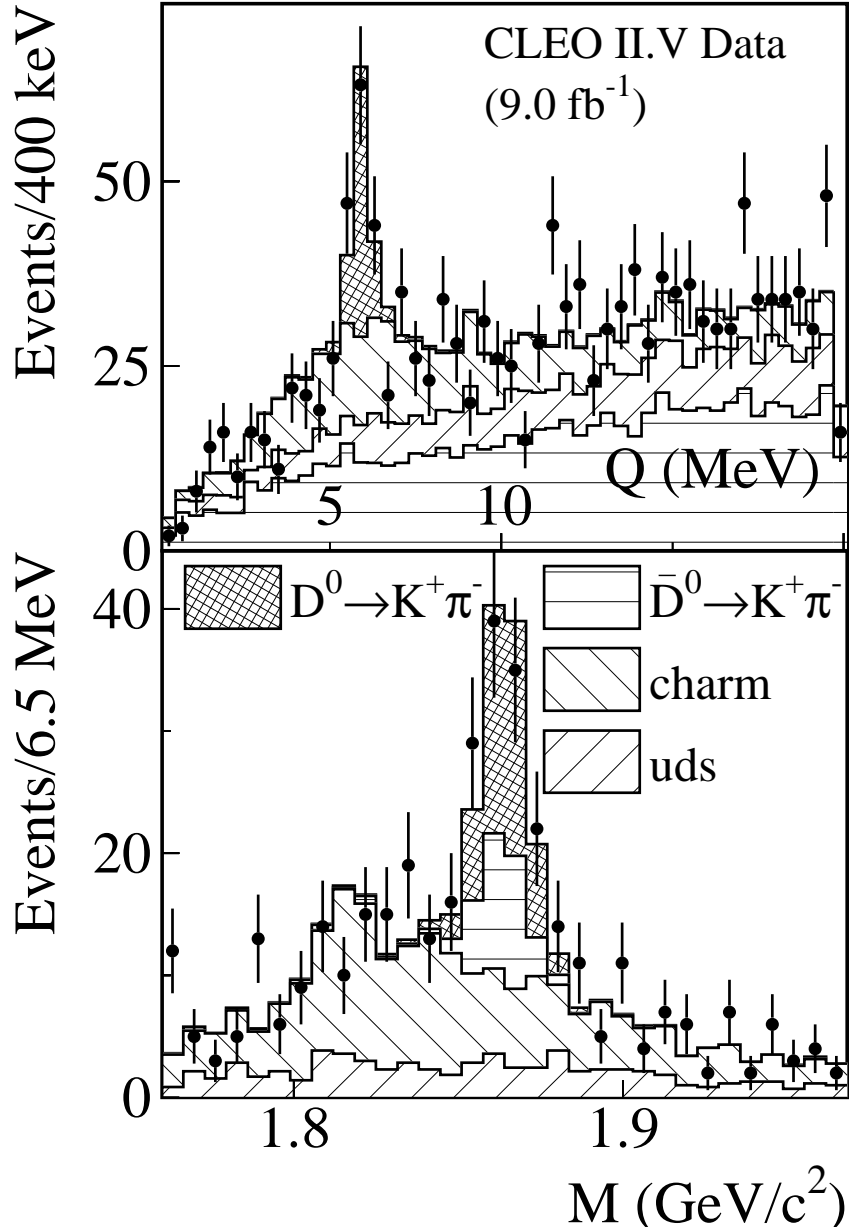}
\end{center}
\caption[Signal for the \hbox{Wrong-Sign} Process $\DZ\!\to\!K^+\pi^-$]
{Signal for the \hbox{wrong-sign} process $\DZ\!\to\!K^+\pi^-$.
For the top plot, $M$ is within $14\,$MeV of the nominal CFD value, and
for the bottom plot, $Q$ is within $500\,$KeV of the nominal CFD value.
The data are the full circles with error bars, the fit to the signal
is cross-hatched, and the fits to the backgrounds are singly hatched.
The results of the fit are summarized in
Table~I.}
\label{fig:qmws}
\end{figure}

Multiple scattering on the field wires, which constitute 70\% of the
material, as measured in radiation lengths, 
in the CLEO-II.V drift chamber, appears to dominate the
current $\sigma_Q$ and $\sigma_M$.  Should the track-fitting be
altered to treat scattering on the field wires as discrete and
localized, both $\sigma_Q$ and $\sigma_M$ might improve by as much
as a factor of 2.

Our signal for the wrong-sign process $\DZ\!\to\!K^+\pi^-$ is shown
in Fig.~\ref{fig:qmws}.  We determine the background levels by performing
a fit to the plane of $0\!<\!Q\!<\!20\,$MeV versus
$1.76\!<\!M\!<\!1.97\,$GeV, which has an area about 150 times larger
than our signal region.  Event samples generated by the Monte-Carlo
method and fully simulated in our detector, corresponding to $90\,$fb$^{-1}$
of integrated luminosity, are used to estimate the background shapes
in the $Q\!-\!M$ plane.  The shapes are allowed to float in a fit to
the data; the results of the fit are displayed in 
Fig~\ref{fig:qmws}.  The excess of events in the signal region is prominent.

We describe the signal shape with the right-sign data that is within
$7\,\sigma$ of the nominal CFD value in the $Q\!-\!M$ plane.  The results
of the fit to the wrong-sign data are summarized in Table~I.

No acceptance corrections are needed to compute
$R_{\rm ws}\!=\!(0.34\pm0.07)\%$ from Table~I.  The dominant
systematic errors all stem from the potentially inaccurate modeling
of the initial and acceptance-corrected shapes of the background
contributions in the $Q$-$M$ plane.  We assess these systematic
errors by substantial variation of the fit regions, $dE/dx$ criteria,
and kinematic criteria; the total systematic error we assess is
$0.06\%$.  

Our complete result for $R_{\rm ws}$ is summarized in Table~II.  As
we describe later, our data are consistent with an absence of
$\DZ\!-\!\DZB$ mixing, so our best estimate for the relative DCSD
rate, $R_D$, is that it equals $R_{\rm ws}$.

There are two directly comparable measurements of $R_{\rm ws}$: one
is from CLEO-II~\cite{tl}, $R_{\rm ws}=(0.77\pm0.25\pm0.25)\%$ which used
a data set independent of that used here; the second is
from Aleph~\cite{alep},
$R_{\rm ws}=(1.84\pm0.59\pm0.34)\%$; comparison of our result and these
are marginally consistent with $\chi^2=6.0$ for 2 DoF, for a C.L. of 5.0\%.

We have split our sample into candidates for $\DZ\!\to\!K^+\pi^-$
and $\DZB\!\to\!K^-\pi^+$.  There is no evidence for a CP-violating
time-integrated asymmetry.  From Table~I, it is straightforward 
to estimate the $1\sigma$ statistical error on the CP violating 
time-integrated asymmetry as $\sqrt{107}/54.8\approx20\%$.
\begin{center}
\begin{minipage}[t]{31.8pc}
TABLE I. Event yields in a signal region of $2.4\,\sigma$
centered on the nominal $Q$ and $M$ values. The
total number of candidates is 107.
The bottom row describes the normalization sample.
\begin{center}
\begin{tabular}{cc}
\textbf{Component}            & \textbf{\# Events} \\ \hline
$\DZ\!\to\!K^+\pi^-$ (WS Signal) & $54.8\pm10.8$ \\
random $\pi^{\pm}+\DZ/\DZB$   & $24.3\pm1.8$ \\
$c\overline{c}$               & $12.3\pm0.8$ \\
$uds$                         & $8.6\pm0.4$ \\ 
$\DZ\!\to\,$Pseudoscalar-Vector  & $7.0\pm0.4$ \\ \hline
$\DZB\!\to\!K^+\pi^-$ (RS Normalization)\hspace{0.5cm} & $16126\pm126$ \\ \hline
\end{tabular}
\end{center}
\end{minipage}
\end{center}
\begin{center}
\begin{minipage}[t]{5.5in}
TABLE II. Result for $R_{\rm ws}$.  For the branching ratio
${\cal B}(\DZ\!\to\!K^+\pi^-)$ we take the absolute branching
ratio ${\cal B}(\DZB\!\to\!K^+\pi^-)=(3.85\pm0.09)\%$. The
third error results from the uncertainty in this absolute branching ratio.
As discussed in the text, our best estimate is that $R_D=R_{\rm ws}$.
\begin{center}
\begin{tabular}{cl}
\textbf{Quantity}            & \textbf{Result} \\ \hline
$R_{\rm ws}$ & $(0.34\pm0.07\pm0.06)\%$ \\
$R_{\rm ws}/\tan^4\theta_C$ & $(1.28\pm0.25\pm0.21)$ \\
${\cal B}(\DZ\!\to\!K^+\pi^-)$\hspace{0.5cm} & $(1.31\pm0.26\pm0.22\pm0.03)\!\times\!10^{-4}$ \\ \hline
\end{tabular}
\end{center}
\end{minipage}
\end{center}

Given the absence of a significant time-integrated CP asymmetry,
we undertake a study of the decay time dependence 
\hbox{wrong-sign} rate based upon Eqn.~\ref{eq:rws}.
We reconstruct the proper $\DZ$ decay time primarily from the
vertical displacement of the $K^+\pi^-$ vertex from the $e^+e^-$
collision `ribbon,' which is infinitesimal in its vertical extent.
We require a well-reconstructed vertex in 3-dimensions,
which causes a loss of about 12\% of the candidates described in Table~I.
Our resolution, in units of the mean $\DZ$ life, is about 1/2.
Study of the plentiful right-sign sample allows us to fix our
detailed resolution function, and shows that biases 
in the reconstruction of the proper decay time contribute negligibly
to the wrong-sign results.

\begin{figure}[htpb]
\begin{center}
\epsfig{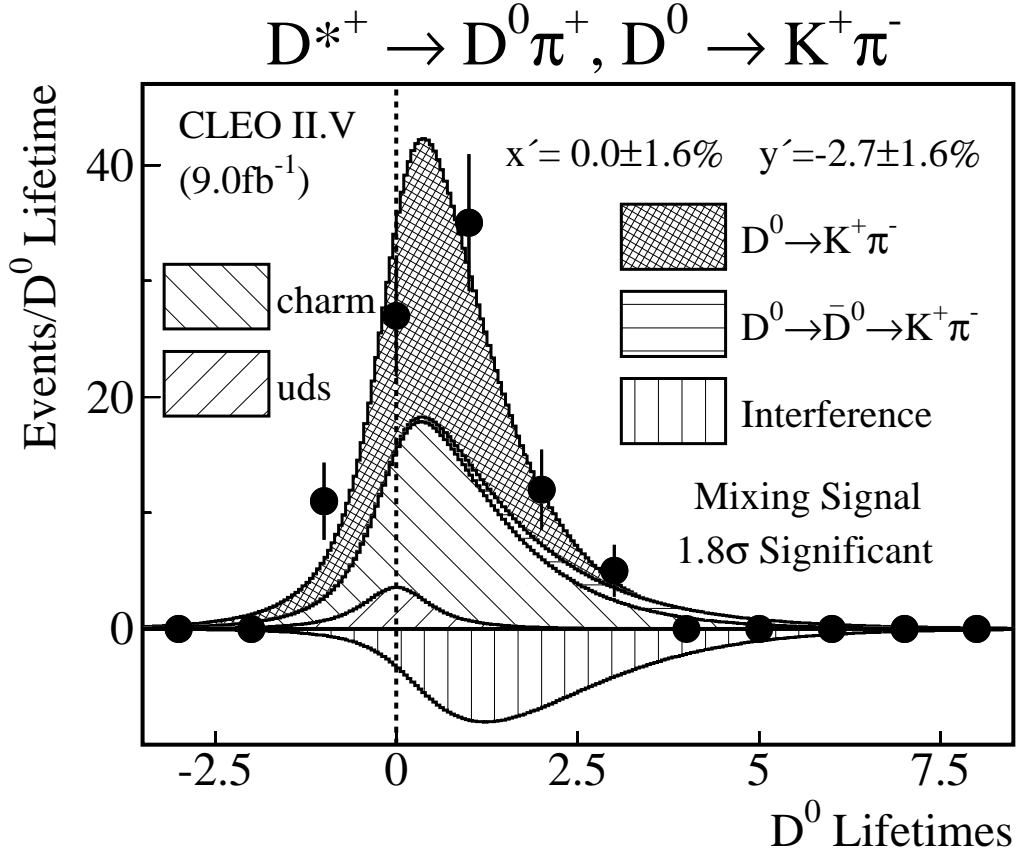}
\end{center}
\caption[Distribution in $t$ for $\DZ\!\to\!K^+\pi^-$.]
{Distribution in $t$ for the $\DZ\!\to\!K^+\pi^-$ candidates.
The data are shown as the full circles with error bars.  All other
information comes from the fit to the data,
The smooth curves show the various specific
contributions as labeled.  The cross hatched region is the net contribution
from $\DZ\!\to\!K^+\pi^-$, after incorporation of the (destructive)
interference and mixing.}
\label{fig:lws}
\end{figure}

The distribution of proper decay times, $t$, for wrong-sign
candidates that are within $2.4\,\sigma$ of the nominal CFD value
in the $Q\!-\!M$ plane is shown in Fig.~\ref{fig:lws}.  Maximum-likelihood
fits are made to those data.  The backgrounds are described
by levels and shapes deduced from the fit to the $Q\!-\!M$ plane,
and from study of the simulated data sample.  The wrong-sign signal
is described by Eqn.~\ref{eq:rws}, folded with our resolution function.

For the benchmark fit to the wrong-sign data,
$x^{\prime}$ and $y^{\prime}$ are constrained
to be zero.  This fit has a
confidence level of $84\%$, indicating a good fit.

The mixing amplitudes $x^{\prime}$ and $y^{\prime}$ are then
allowed to freely vary, and the best fit values are shown in
Fig.~\ref{fig:lws} and in Table~III.
The fit improves slightly when $x^{\prime}$ and $y^{\prime}$ are
allowed to float to the values that maximize the likelihood.
However, the small value of the likelihood change
$\sqrt{-2\Delta\ln{\cal L}}=1.8\,\sigma$ (including
systematic errors) does not permit us to eliminate the possibility
that the improvement is due to a statistical fluctuation.

Therefore, our principal results concerning mixing are the
one-dimensional intervals, which correspond to a 95\% confidence level,
that are given in the second column of Table~III.

\begin{figure}[htpb]
\begin{center}
\epsfig{figure=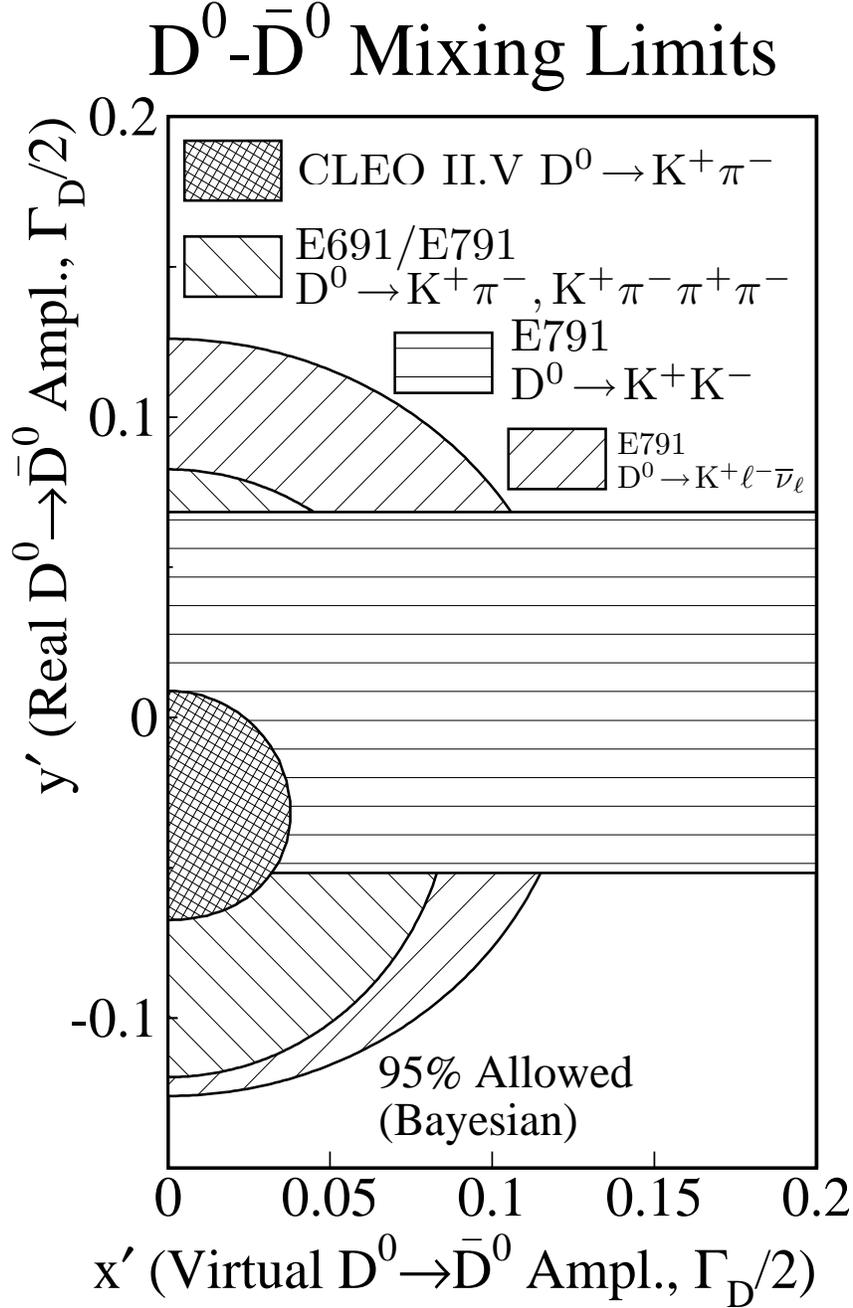,width=26.5pc}
\end{center}
\caption[Limits in the $y^\prime$ vs. $x^\prime$ Plane]
{Limits in the $y^\prime$ vs. $x^\prime$ plane.  Our experiment
limits, at 95\% C.L., the true values of $x^\prime$ and
$y^\prime$ to occupy the cross-hatched  region near the origin.
Also shown are the similar zones from other recent experiments.
We assume $\delta=0$ to place the recent work of E791
that utilized $\DZ\!\to\!K^+K^-$;
a non-zero $\delta$ would
rotate the E791 confidence region clockwise about the origin
by an angle of $\delta$.}
\label{fig:xyl}
\end{figure}

Additionally, we evaluate a contour in the two-dimensional
plane of $y^{\prime}$ versus $x^{\prime}$ which, at
95\% confidence level, contains the true value of $x^{\prime}$ and
$y^{\prime}$. To do so, we determine
the contour around our best fit values where the $-\ln{\cal L}$ has increased
by 3.0~units.  All other fit variables, including the DCSD rate and
background contributions, are allowed to float to distinct, best fit
values at each point on the contour.
The interior of the contour is shown, as the small,
dark, cross-hatched region near the origin of Fig.~\ref{fig:xyl}.
On the axes of $x^\prime$ and $y^\prime$, this contour
falls slightly outside the one-dimensional intervals listed in Table~III,
as expected.\vspace{-3ex}
\begin{center}
\begin{minipage}[t]{5.5in}
TABLE~III. Results of the fit, where $x^{\prime}$ and
$y^{\prime}$ are free to float, to the  
distribution of $\DZ\!\to\!K^+\pi^-$ candidates in $t$.  The data and
the fit components are shown in Fig.~\ref{fig:lws}.
\begin{center}
\begin{tabular}{ccc}
\textbf{Parameter} & \textbf{Best Fit} & \textbf{95\% C.L.}\\ \hline
\vphantom{$A^{A^A}$}$R_D$      & $(0.50^{+0.11}_{-0.12}\pm0.08)\%$ & 
$0.22\%<R_D<0.77\%$\\
$y^\prime$ & $(-2.7^{+1.5}_{-1.6}\pm0.2)\%$ & $-5.9\%<y^\prime<0.3\%$ \\ \hline
$x^\prime$ & $(0\pm1.6\pm0.2)\%$             & $|x^\prime|<3.2\%$ \\
$(1/2)x^{\prime2}$ &     & $<0.05\%$ \\ \hline
\end{tabular}
\end{center}
\end{minipage}
\end{center}

We have evaluated the allowed regions of other experiments
\cite{E791,E691,E791L,E791CP} at 95\% C.L., and shown those
regions in Fig.~\ref{fig:xyl}. 

All results described here are preliminary.

If we assume that $\delta$ is small, which
is plausible~\cite{wfsp,brpa} then $x^\prime\!\approx\!x$
and we can indicate the impact of our work in limiting predictions
of $\DZ\!-\!\DZB$ mixing from extensions to the Standard Model.
Eighteen of the predictions recently tabulated~\cite{hnncomp} have
some inconsistency with our limit.  Among those predictions,
some authors have made common assumptions, however.

Because our data is consistent with an absence of $\DZ\!-\!\DZB$ mixing,
our best information on the DCSD rate, $R_D$, is that it equals
$R_{\rm ws}$, as summarized in Table~II.

We will soon complete studies where we allow various types of
CP violation to modify Eqn.~\ref{eq:rws}.  Also,
the $\DZ$ decay modes $K^+K^-$, $\pi^+\pi^-$, $\KSHO\phi$, $\KSHO\pi^+\pi^-$, 
$K^+\pi^-\pi^0$, $K^+\pi^-\pi^+\pi^-$, and $K^+\ell^-\overline{\nu}_l$
are under active investigation, using the CLEO-II.V data.
The $(\KSHO\pi^+)$ resonance bands in
$\DZ\!\to\!\KSHO\pi^+\pi^-$ permit investigation of $\DZ\!\to\!\DZB$, with
sensitivity heightened by coherent interference that is modulated
by resonant phases.

The entire CLEO-II.V data set, suitably exploited, could be used to
observe $\DZ\!-\!\DZB$ mixing if either $|x|$ or $|y|$ exceed approximately
$1\,\%$.

I gratefully acknowledge the efforts of the CLEO Collaboration,
the CESR staff, and the staff members of CLEO institutions.  
This work is David Asner's Ph.D. Dissertation topic, and he
made the principal contributions to this study, and
Tony Hill has made important contributions.
Rolly Morrison and Mike Witherell provided
intellectual guidance.

I thank the hosts of Kaon `99, particularly Y.~Wah and B.~Winstein,
for running a hospitable and stimulating conference.  I thank J.~Rosner
for help with this manuscript.

This work was supported by the Department of Energy under contract
DE-AC03-76SF00098.


\begin{thebibliography}{99}
\bibitem{xyk}Specifically, from the previous reference, for the
$\KZ$ system $x\!=\!-0.947\pm0.003$, and $y\!=\!0.99655\pm0.00003$.  In our
convention $|y|$ is bounded by $1$, and signed by the CP eigenvalue of the
dominant real intermediate state(s), which is $2\pi$ (to which we assign
CP=+1) for $\KZ\!\to\!\KZB$.  For $\DZ\!\to\!\DZB$ there are real
intermediate states that are not CP eigenstates, such as $K\pi$, but
the convention is straightforward
to generalize.  Then, we define $x$ as proportional
to the mass difference between the eigenstate that decays preferentially
to CP=$+1$ final states and the other eigenstate, that decays preferentially
to CP=$-1$ states.  The study reported here is sensitive 
(when $\delta=0$) to
the magnitude $|x|$ for the $\DZ\!\to\!\DZB$ amplitude and to the
magnitude, and sign, of $y$ for the $\DZ\!\to\!\DZB$ amplitude.
\end{thebibliography}
\end{document}